
%
\let\useblackboard=\iftrue
%
%
\newfam\black
\input harvmac.tex
\let\includefigures=\iffalse
\let\expandedversion=\iffalse
\includefigures
\message{If you do not have epsf.tex (to include figures),}
\message{change the option at the top of the tex file.}
\input epsf
\epsfclipon
\def\fig#1#2{\topinsert\epsffile{#1}\noindent{#2}\endinsert}
\def\ebox#1#2{\topinsert\epsfbox{#1}\noindent{#2}\endinsert}
\else
\def\fig#1#2{\vskip .5in
\centerline{Figure 1}
\vskip .5in}
\def\ebox#1#2{\vskip .5in
\centerline{Figure 2}
\vskip .5in}
\fi
\def\Title#1#2{\rightline{#1}
\ifx\answ\bigans\nopagenumbers\pageno0\vskip1in%
\baselineskip 15pt plus 1pt minus 1pt
\else
\def\listrefs{\footatend\vskip 1in\immediate\closeout\rfile\writestoppt
\baselineskip=14pt\centerline{{\bf References}}\bigskip{\frenchspacing%
\parindent=20pt\escapechar=` \input
refs.tmp\vfill\eject}\nonfrenchspacing}
\pageno1\vskip.8in\fi \centerline{\titlefont #2}\vskip .5in}

\ifx\answ\bigans\def\tcbreak#1{}\else\def\tcbreak#1{\cr&{#1}}\fi
\useblackboard
\message{If you do not have msbm (blackboard bold) fonts,}
\message{change the option at the top of the tex file.}
\font\blackboard=msbm10 scaled \magstep1
\font\blackboards=msbm7
\font\blackboardss=msbm5
\textfont\black=\blackboard
\scriptfont\black=\blackboards
\scriptscriptfont\black=\blackboardss
\def\Bbb#1{{\fam\black\relax#1}}
\else
\def\Bbb{\bf}
\fi
%
\def\yboxit#1#2{\vbox{\hrule height #1 \hbox{\vrule width #1
\vbox{#2}\vrule width #1 }\hrule height #1 }}
\def\fillbox#1{\hbox to #1{\vbox to #1{\vfil}\hfil}}
\def\ybox{{\lower 1.3pt \yboxit{0.4pt}{\fillbox{8pt}}\hskip-0.2pt}}
\def\nl{\hfill\break}

\def\comments#1{}

\def\BR{\Bbb{R}}

\def\p{\partial}

\def\half{{1\over 2}}

\def\tr{{\rm tr\ }}
\def\Str{{\rm Str\ }}

\def\CN{{\cal N}}

\def\I{I}
\def\II{II}
\def\IIa{IIa}
\def\IIb{IIb}

\Title{\vbox{\baselineskip12pt
\hfill{\vbox{
\hbox{hep-th/9512077 ~~~ RU-95-92\hfil}
}}}}
{\vbox{\centerline{Branes within Branes}
}}
\centerline{Michael R. Douglas}
\smallskip
\centerline{Dept. of Physics and Astronomy}
\centerline{Rutgers University }
\centerline{Piscataway, NJ 08855-0849}
\centerline{\tt mrd@physics.rutgers.edu}
\bigskip
\bigskip
\noindent
We discuss a set of universal couplings between superstring
Ramond-Ramond gauge fields and the gauge fields internal to D-branes,
with emphasis on their topological consequences,
and argue that instanton solutions in these internal theories are
equivalent to D-branes.
A particular example is the Dirichlet $5$-brane in type \I\ theory,
which Witten recently showed is is the zero size limit of an
$SO(32)$ instanton.
Its effective world-volume theory is an $Sp(1)$ gauge theory,
unbroken in the zero size limit.
We show that the zero size limit of an instanton in this theory
is a $1$-brane, which can be described as a bound state of
the Dirichlet $1$-brane with the $5$-brane.
Considering several $1$ and $5$-branes provides
a description of moduli spaces of $Sp(N)$ instantons,
and a type \II\ generalization is given which should describe
$U(N)$ instantons.

\Date{December 1995}
\nref\dual{J.~H.~Schwarz and A.~Sen, Phys. Lett. {\bf B312} (1993) 105;\nl
C. Hull and P. Townsend, Nucl. Phys. {\bf B438} (1995) 109;\nl
E. Witten, Nucl. Phys. {\bf B443} (1995) 85;\nl
A. Strominger, Nucl. Phys. {\bf B451} (1995) 96.}
\nref\dlp{J.~Dai, R.~G.~Leigh and J.~Polchinski, Mod. Phys. Lett. {\bf A4}
(1989) 2073.}
\nref\pol{J.~Polchinski, ``Dirichlet branes and Ramond-Ramond charges,''
hep-th/9510017.}
\nref\leigh{R.~G.~Leigh, Mod. Phys. Lett {\bf A4} (1989) 2767.}
\nref\green{M.~B.~Green, Nucl. Phys. {\bf B293} (1987) 593-611.}
\nref\gsw{M.~B.~Green, J.~Schwarz and E.~Witten, Superstring Theory vol. \II,
section 14.3.4, pp.387-390.}
\nref\pc{J.~Polchinski and Y.~Cai, Nucl.~Phys. {\bf B296} (1988) 91.}
\nref\callan{C. G. Callan, C. Lovelace, C. R. Nappi and S. A. Yost,
Nucl. Phys. {\bf B308} (1988) 221.}
\nref\chs{C. Callan, J. Harvey and A. Strominger, Nucl. Phys.
{\bf B367} (1991) 60.}
\nref\witadhm{E. Witten, ``Sigma Models and the ADHM Construction of
Instantons,'' hep-th/9410052.}
\nref\witbound{E. Witten, ``Bound States of Strings and p-branes,''
hep-th/9510135.}
\nref\li{M.~Li, ``Boundary States of D-Branes and Dy-Strings,''
hep-th/9510161.}
\nref\lidil{M.~Li, ``Dirichlet Boundary State in Linear Dilaton Background,''
hep-th/9512042.}
\nref\senu{A.~Sen, ``U-duality and Intersecting D-branes,'' hep-th/9511026.}
\nref\ck{C. G. Callan and I. R. Klebanov, ``D-Brane Boundary State
Dynamics,'' hep-th/9511173.}
\nref\bachas{C.~Bachas, ``D-brane Dynamics,'' hep-th/9511043}
\nref\polwit{J.~Polchinski and E.~Witten, ``Evidence for Heterotic-Type I
String Duality,'' hep-th/9510169.}
\nref\horava{P.~Horava and E. Witten, ``Heterotic and Type I String Dynamics
from Eleven Dimensions,'' hep-th/9510209.}
\nref\witten{E. Witten, ``Small Instantons in String Theory,'' hep-th/9511030.}
\nref\morrison{D.~Morrison, ``Mirror Symmetry and the Type II String,''
hep-th/9512016.}
\nref\nepteit{R.~Nepomechie, Phys. Rev. {\bf D31} (1985) 1921;\nl
C.~Teitelboim, Phys. Lett. {\bf B167} (1986) 63, 69.}
\nref\crem{E. Cremmer and J. Scherk, Nucl. Phys. {\bf B72} (1974) 117.}
\nref\christ{N.~H.~Christ, E.~J.~Weinberg and N.~K.~Stanton, Phys. Rev.
{\bf D18} (1978) 2013-2025.}
%

Superstring theory has a rich spectrum of solitonic states,
and over the last years
much has been learned about their important roles in the theory,
in strong-weak coupling duality and in resolving singularities.~\dual\ %
The most important are the BPS states (those in reduced multiplets of
supersymmetry) as one can make exact statements about them even at
strong coupling, and thus dualities can predict their existence.
Many dualities exchange elementary string states with states having
Ramond-Ramond gauge charge, which strongly suggests that the two should be
considered equally fundamental.

In the remarkable work \pol\ Polchinski showed that the Dirichlet branes of the
Type \I\ and \II\ superstrings are RR charged BPS states.
A Dirichlet (D) brane~\refs{\dlp} is simply an allowed endpoint for open
strings and these
objects are much more tractable than previously studied string solitons.
At low energies, a D-brane is described by an
effective world-volume theory, essentially
dimensionally reduced supersymmetric Yang-Mills theory.~\refs{\leigh,\witbound}
The difficulty of the non-renormalizability of $p$-brane theories for $p>1$
is resolved: these theories are the low-energy limits of UV-finite open string
theories. (Clearly it will be very interesting to elucidate this point.  An
old proposal of Green (for $p=1$) was in this spirit. \green)

Many of the objects required by string duality have now been identified
as D-branes, such as the heterotic string soliton \polwit.
In another remarkable recent work \witten\
the $5$-brane of type \I\ superstring theory was studied
by Witten, who proposed that it was the zero size limit of a gauge
instanton.
The instanton $5$-branes had been studied as field theory solitons and
as conformal field theories (in the dual heterotic theory).~\chs\ %
The construction starts with a Yang-Mills instanton, which by the anomalous
Bianchi identity $dH=\tr R^2-\tr F^2$ is an object
with RR charge.  The complete solution has metric and dilaton dependence
as well, and in the limit of small instanton scale size a long tube develops
with large string coupling.
Witten's treatment provides a simple description of the resulting
physics: the zero size limit is an
$Sp(1)$ gauge theory in the world-volume.

This world-volume theory can have non-perturbative physics of its own,
and in this note we study the properties of its instanton solution,
a $1$-brane confined to the $5$-brane,
and show that it can be described as a bound state of the Dirichlet
$1$-brane with the $5$-brane.  We also mention type \II\ generalizations
to $p-4$ branes within $p$-branes.

\medskip
\def\hH{{\hat H}}
We begin by reviewing the RR fields of
superstring theory and their generalized Green-Schwarz couplings.
The massless RR gauge fields are naturally combined into a bispinor
potential $C_{\alpha\tilde\beta}(x)$ and field strength
$H_{\alpha\tilde\beta}$ in the
$(16+16')\otimes (16+16')$
Dirac spinor representations of the algebra of
fermion zero modes $\psi_0^\mu$ and $\tilde\psi_0^\mu$.
This is equivalent to a sum of differential forms, e.g.
$C = \sum C^{(p+1)}_{\mu_1\ldots\mu_{p+1}}
\Gamma^{(p+1)}_{\mu_1\ldots\mu_{p+1}}$,
with $\delta_{\alpha\tilde\beta}$ as the $0$-form and using 
$b^{*\mu} \equiv dx^\mu \equiv  \half(\psi_0^\mu+i\tilde\psi_0^\mu)$
and
$b^{\mu} \equiv *dx^\mu* \equiv  \half(\psi_0^\mu-i\tilde\psi_0^\mu)$.
The on-shell conditions for the RR vertex operator
$\hH_{\alpha\tilde\beta} S^\alpha\tilde S^{\tilde\beta}~\exp ikX$ are
separate Dirac equations $b^{*\mu}\p_\mu \hH = b^{\mu}\p_\mu \hH = 0$,
whose sum and difference are
$d\hH = *d*\hH = 0$.
Thus $\hH$ is an RR field strength and the free equation of motion
and Bianchi identity appear on the same footing.
The GSO projections $(-1)^{F+\tilde F}H=(-1)^{\tilde F}H=H$
will restrict $H$ to contain only even (for \IIa)
or odd (\I\ and \IIb) forms, and impose $\hH=*\hH$.
In type \I\ theory we also impose $\Omega \hH = \hH$,
where the twist operator $\Omega$ acts as
$H\rightarrow i^{r-3}H$, keeping only $H^{(3)}$
and its dual $H^{(7)}$.

The tension and RR charge of a $D$-brane can be determined from one-point
functions on a disk with boundary on the
$D$-brane.  By usual string coupling counting these would be
$O(1/\lambda)$, correct for the tension but not the charge.
This is because
the conserved RR charges are not surface integrals of $\hH$.
A simple way to see this is to add
a linear dilaton background $D\propto X$.~\pol\ %
The world-sheet supercurrents become
$G=\psi\cdot\p X + (\p D)\cdot\p\psi$ and the Bianchi identity
becomes
\eqn\formeq{d\hat H - \p_\mu D dx^\mu \hat H = e^{-D} d e^{-D} \hat H = 0.}
(We also incorporated the usual $e^{-2D}=1/\lambda_{str}^2$ for an inverse
propagator.  This is a bit cavalier but will suffice for our purpose;
see \lidil\ for a better treatment.)
Thus the RR charges are surface integrals of
$H \equiv e^{-D} \hat H \equiv dC$.

A $D$-brane acts as an electric $p+1$-form source in \formeq,
\eqn\formeqs{d H = {\mu_p\over\alpha} *J^{(p+1)}}
where $J^{(p+1)}$ is the $p+1$-form tangent to the brane.
$\mu_p$ is the charge and the RR action is
$\sum_{p\le 3}\half\alpha\int H^{p+2}*H^{p+2}$.
(The common origin of the RR fields makes such a convention natural.
We will take $\alpha'=1$, $\alpha=1/2\pi$, and
$\mu_p=(2\pi)^{3-p}$).
This was inferred from the closed string exchange in \pol;
it can also be computed along the lines of the open string tadpole computation
in \refs{\pc,\callan}.  The result in the massless sector can be derived
from the result $\delta_{\alpha\tilde\beta}$ for a free boundary ($p=9$)
by changing to Dirichlet boundary conditions in the orthogonal directions,
which takes $\tilde\psi^\mu\rightarrow-\tilde\psi^\mu$.
The allowed values of $p$ in a particular string theory
are those for which the appropriate form exists,
so in type \I\ theory $1$, $5$ and $9$-branes exist.

The result \formeqs\ was generalized to
a constant background gauge field
by Li \li, using the open string results of \refs{\callan,\pc}.
The zero momentum limit in the massless sector is particularly simple and
can be derived simply by letting the boundary action for the gauge field
$\exp \int d\sigma\ F_{\mu\nu} \psi^\mu\psi^\nu$
act on the boundary state.
Since the gauge field vertex operators have $\mu,\nu$ tangent to the
brane they are just $F_{\mu\nu} b^{*\mu} b^{*\nu}=F_{\mu\nu} dx^{\mu} dx^{\nu}$
and the result is
\eqn\gsource{dH = {\mu_p\over\alpha} \Str e^{F} *J^{p+1}.}
As argued in \callan\ this generalizes to the non-abelian case with
$F=dA+A^2$.
This implies a world-volume action
\eqn\sym{S =  \int
T_p e^{-D}\sqrt{\det (h + F)} + \mu_p\ C\wedge \Str e^{F}. }
The $9$-brane result includes both the $dH=\tr F^2$
term in the anomalous Bianchi identity and the $C^{(2)} \tr F^4$ coupling
required by the Green-Schwarz mechanism.
The generalization shows the
close connection of these with the field theory gauge anomaly.

Using $T$-duality, it is just as easy to put in the dependence on the
transverse modes $X^i$ of a general $p$-brane:
\eqn\gsourcetwo{dH = {\mu_p\over\alpha}\Str e^{F + dX^i b_i} *J^{p+1}.}
One role of this term is to implement Lorentz covariance:
if we take $X^i = R^i_\alpha x^\alpha$, we get a rotated or boosted
$D$-brane, and the boundary state is proportional to
$\prod d(X^i+R^i_\alpha x^\alpha)$.
In the spirit of Bachas \bachas, we could also
argue in the other direction: the generalized Green-Schwarz coupling
(and by extension the field theoretic gauge anomaly)
follows simply from $T$-duality and Lorentz invariance!

In type \II\ string theories, there is a $U(1)$ factor in the D-brane
gauge symmetry, which will mix with the NS $2$-form tensor
to preserve its gauge invariance  $\delta B=-d\Lambda$,
modifying \gsource\ as $F\rightarrow F-B$.~\refs{\crem,\witbound}\ %
The $B^k C$ couplings produced by this should be those predicted by
Morrison from considerations of mirror symmetry. \morrison

Another test of \gsource\ is that localized gauge field solutions
in a $D$-brane should be sources with correctly quantized charge.
Now terms in the expansion of $\Str\exp\ iF/2\pi$  are the Chern characters
(which appear in the index theorem) which will indeed integrate
to integers on localized gauge fields such as instantons.
The $2\pi$'s needed are
reproduced by the dependence of the $\mu_p$ on $p$.

There is a subtlety in that we must insist on the word `localized'
in the above statement.
For example, the dyonic $1$-branes of \IIb\ theory are sources of $C^0$ with
charge $\lambda m$, which is not quantized in the unit set by the
D-instanton charge.  Now there is no paradox yet as the Dirac argument
\nepteit\ does not
apply to a zero-form potential, but clearly the same phenomenon could make
higher branes non-quantized sources of the higher potentials.
As a test case, consider a flux tube,
say $F_{01}\ne 0$ in a $5$-brane, which we take to satisfy Gauss' law by simply
requiring $\p_0 F_{01} = \p_1 F_{01} = 0$.
This will be a source of $C^{4}$, but only components with an index in the
$01$ dimensions, for which the Dirac argument again does not go through.

Because of \gsource, an instanton in a $p$-brane will be an
magnetic source of $H^{(12-p)}$
(or electric source of $H^{(p-2)}$).
This generalizes the result of \chs, for the solitonic $5$-brane constructed
from a type \I\ instanton ($p=9$), to general $p$.
It motivates a generalization of Witten's proposal:
the zero scale size limit of the instanton in a $p$-brane will be
a Dirichlet $p-4$ brane, bound to the $p$-brane.
We will study this in detail for $p=5$:
the instanton in Witten's $5$-brane, which on general grounds is a $1$-brane,
will turn out in the zero size limit
to become a bound state of the Dirichlet $1$-brane with the
Dirichlet $5$-brane.

\medskip
This brings us to the structure of the $1$ and $5$-branes
of type \I\ theory, analyzed by Polchinski and
Witten.~\refs{\polwit,\witten}\ %
First consider the Dirichlet $1$-brane, which will have electric $C^{(2)}$
charge.  This and its string tension $T\sim 1/\lambda$ make it a candidate
for the string soliton required by type \I-heterotic duality.
The first point to check is its supersymmetry, which is half that
of the type \I\ theory: in other words $16 = 8^+ + 8'^-$ under
$SO(8)\times SO(1,1)$ of which the first part is unbroken, so the
world-sheet theory will have a chiral $(0,8)$ supersymmetry.

To make a more detailed comparison of the world-sheet theories,
we must realize that there are two important
modifications from the SYM theory \sym.
First, we have an additional projection on the world-sheet theory,
$\Omega=1$.
With a single $1$-brane this produces $SO(1)$ gauge theory, which has no
massless gauge sector.  On the other hand, $\Omega$
acts with opposite sign on the vertex operators $\p_\tau X$ and
$\p_\sigma X$, so the transverse coordinates $X$ survive the projection.
The Ramond sector contains their superpartners $\psi$, a spinor $8$ of $SO(8)$.
These are chiral on the world-sheet, both because the supersymmetry is
chiral, but as can also be seen by considering the action of $\Omega$.

Second, we have an additional sector of open strings
(the `DN sector')
with mixed boundary
conditions, one on the $1$-brane and one free (i.e. on a $9$-brane).
A free superfield with mixed Dirichlet-Neumann boundary conditions has
moding shifted by $1/2$ and ground state energy shifted by $1/8$, so the
NS sector has ground state energy $-1/2+8/8>0$ and no massless states,
while the R sector is a $SO(1,1)$ spinor $\lambda^I$ in the $32$ of $SO(32)$,
chiral after the GSO projection.  We thus reproduce exactly the zero
modes of the light-cone heterotic string.

We turn to Dirichlet $5$-branes.
One such object can be obtained by $T$-duality from the $9$-brane,
but Witten's $5$-brane~\witten\ is different: it has $Sp(1)$
Chan-Paton factors.  In other words, the open strings ending on the
$5$-brane have a two-valued index $A$, and the twist $\Omega$ acts
on this index as $T:v_A\rightarrow \epsilon_{AA'} v_{A'}$.
The opposite sign for the displacement vertex operator makes it
a two-index antisymmetric tensor of $Sp(1)$, i.e. a scalar, so this also
describes a single object, now with charge $2$.
This charge is consistent with the normalizations in \gsource\ if we
remember that a minimally embedded $SO(N)$ instanton has charge
$1/8\pi^2 \int~\tr F^2= 2$.

The type \I\ supersymmetry now decomposes under $SO(5,1)\times SO(4)$
as $16=(4,2)+(4',2')$.  Taking the first half as unbroken, we get
a single Weyl $d=6$ spinor, so the world-volume theory will contain
$N=1$, $d=6$ $Sp(1)$ SYM theory and a singlet hypermultiplet $X$.
The $\Omega$ projection acts on the gauginos as
$\Gamma^1\Gamma^2\Gamma^3\Gamma^4 T$, correlating their symmetry as
$Sp(1)$ tensors with their $d=6$ chirality, as required by supersymmetry.

The DN sector is a $(2,32)$ of $Sp(1)\times SO(32)$.
The NS sector has ground state energy $-1/2+4/8=0$ so contributes
six dimensional scalars, which evidently must be in hypermultiplets.
Indeed the fermion zero modes $\psi_0^i$ make these states a spinor
$(1,2)$ of $SO(4)$ (after GSO).
Since the two ends of the string are different,
the $\Omega$ projection relates these fields with their charge conjugates.
It includes an $\epsilon_{AB}$ on the Chan-Paton factors,
and an $\epsilon_{\alpha\beta}$ to undo the $SO(4)$ charge conjugation,
leaving `half-hypermultiplets' whose scalars satisfy
\eqn\halfhy{
H^*_{\alpha AI}=\epsilon_{AB}\epsilon_{\alpha\beta}H_{\beta BI}.}
The R sector is a chiral $d=6$ spinor with
$\chi^{AI*}=C\chi^{BI}\epsilon_{AB}$.

The scalars $\phi_{\alpha AI}$ provide flat directions which
break $SO(32)$.  Indeed Witten showed that this moduli space coincides
precisely with the ADHM description of instanton moduli space.
It is the space of solutions of the $D$-flatness
conditions (equivalent to the $D$ and $F$ conditions of the more familiar
$N=2$, $d=4$ case)
\eqn\dflat{
0=\sum_I H^*_{\alpha AI} \sigma^i_{\alpha\beta} t^j_{AB}H_{\beta BI}.}
A way to see the content of these conditions is to write the $\phi$ as
$SO(4)$ vectors $H_{\alpha AI}=\sigma^\mu_{\alpha A} H^\mu_I$.
The reality condition \halfhy\ then just states that the $H^\mu_I$
are real.  Since the $9$ $D$-flatness conditions form an $SO(4)$-covariant set,
they must be expressable as $H^\mu\cdot H^\nu=c\delta^{\mu\nu}$,
and the four vectors $H^\mu/\sqrt{c}$ form an orthonormal frame.
Thus the moduli space is a Grassmannian $SO(32)/SO(28)\times SO(4)$
times a $\BR^4$ including scale size and an overall $SU(2)$ gauge rotation.
By considering $k$ coincident $5$-branes,
the construction generalizes to $k$-instanton moduli space.  The effective
theory is $Sp(k)$ gauge theory with matter $X_{[AB]}$ and $H_{\alpha AI}$.


\medskip
We now turn to our main example, and the question:
Now that we know a $5$-brane has $Sp(1)$ gauge symmetry, what are
the properties of a instanton solution of this gauge theory?
It will be a $1$-brane embedded in the $5$-brane, and
it will break half of the $d=6$ supersymmetry.
This decomposes under $SO(4)\times SO(1,1)$ as the complex spinors
$2^+ + 2'^-$ and we see that the world-sheet theory will have $(0,4)$
supersymmetry.
Its tension is proportional to the instanton action in the $5$-brane,
of order $1/\lambda$.
Finally, from \gsource, we expect it to have
the electric $H^{(3)}$ charge of a Dirichlet $1$-brane.

Before exploring the possibility that this is a $D$-brane,
let us work out the field content of the solitonic $1$-brane
of the $5$-brane theory.  Each zero mode of the theory dimensionally
reduced to $N=2$, $d=4$ will provide a world-sheet field.
There is the usual $5$-dimensional moduli space of $Sp(1)$
instantons, position and a scale size, giving transverse fluctuations and
an additional scalar, and we include global
$Sp(1)$ gauge rotations as well for $8$ bosonic zero modes.
The gauginos have $8$ zero modes which are partners under the broken
supersymmetry of these bosonic modes.  
Finally, the fermions in the matter hypermultiplets will provide $32$
zero modes (these are half-flavors of $SU(2)$ in $d=4$), while there are
no covariantly constant scalars in the instanton background.

Both the charge and the counting of zero modes
tell us that if this $1$-brane is a $D$-brane,
it must be a bound state between the $5$-brane and the
$1$-brane we already discussed!
Let us consider a state containing a $1$-brane embedded in the $5$-brane.
The space-time symmetry is now $SO(4)_I \times SO(4)_E \times SO(1,1)$
with the factor `$I$' inside the $5$-brane and the factor `$E$' external
to both.  The $5$-brane supersymmetry will decompose as
$(2,2)^+ + (2',2)^-$, with the first factor unbroken by the one-brane.
Besides the fields $(X,\psi)$ and $\lambda$,
there is an open string sector $DN'$ with strings stretching from the
$1$-brane to the $5$-brane, producing fields in the $2$ of $Sp(1)$.
The NS sector again has ground state energy zero and will provide
world-sheet scalars $Y$ in the $(2,1)$ of $SO(4)_I$.
The R sector has fermion zero modes both in the world-sheet dimensions
and in the $4$ dimensions transverse to both $1$-brane and $5$-brane.
Thus it provides fermions $\rho$ in the $(1,2)^+$
and $\bar\rho$ in the $(1,2')^-$.

The proposal is that an instanton in the $5$-brane is just this object
with the fields $Y$ turned on, and that doing this will make
the components of $(X,\psi)$ perpendicular to the $5$-brane massive,
binding it to the $5$-brane.
The $32$ zero modes are present, while the transverse displacements will
be a subset of $(X,\psi)$ and the gaugino zero modes are $\rho$.
Now the $DN'$ states have masses $m^2\sim (X_1-X_5)^2$
(the displacement of the $1$-brane perpendicular to the $5$-brane),
so there must be a potential term $Y^2(X_1-X_5)^2$ as well as
mass terms $\rho\bar\rho(X_1-X_5)$.
At $X_1=X_5$ the $Y$'s become massless,
and assuming there is no potential for $Y$ (which we will argue below),
we can move out on the branch $Y\ne 0$, giving
a mass to the transverse components of $X_1-X_5$.
We also need their fermionic partners to become massive,
so there must also be couplings $\psi\bar\rho Y$.
In fact all of these couplings are related by $(0,4)$ supersymmetry.

By bringing $k$ $1$-branes and $N$ $5$-branes together, we get an
$SO(k)\times Sp(N)$ gauge theory, whose vacua should parameterize
the moduli space of $k$ instantons in $Sp(N)$ gauge theory.
The matter content is a symmetric tensor $X_{(ij)}$ of $SO(k)$, their
left-moving partners $\psi_{(ij)}$ in the $(2,2)+(2',2')$,
right-moving antisymmetric $\bar\psi_{[ij]}$ in the $(2,2')+(2',2)$
and $\lambda_{iI}$ in the
$(k,1,32)$, scalars $Y$ in the $(k,2N,1)$ and the $2$ of $SO(4)_I$,
and complex fermions in the $(k,2N,1)$ and $(1,2)^+ + (1,2')^-$.
Turning on enough $Y$ to bind all of the $1$-branes, the moduli space
will have the correct dimension $4kN+2k(k+1)-2k(k-1)=4k(N+1)$,
if the potential is the same one we would have in $d=6$, $\CN=1$
$SO(k)$ gauge theory, a point we will argue shortly.
Leaving out the $\lambda$'s (which are obvious zero modes of $\chi$)
there are $4k(k+1)+4kN$ left-moving
and $4k(k-1)+4kN$ right-moving fermions, out of which we need
$4k(N+1)$ left movers to stay massless, $4k^2$ left and right to pair up,
and the remaining $4k(N-1)$ right movers must correspond to additional
zero modes in a $k$-instanton background of the $5$-brane theory.
Indeed this is the right number to come from the antisymmetric tensor.

We needed the potential for $Y$ to be that of a $d=6$, $\CN=1$
supersymmetric gauge theory,
but the $1$-brane is a $d=2$, $(0,4)$ theory.  We argue
as follows: the DN vertex operators for $Y$ are twist fields in the four
dimensions transverse to the $1$-brane and internal to the $5$-brane, and
trivial in the other six dimensions.
In fact they are exactly the same DN vertex operators as in Witten's
$5$-brane, there twist operators in the
four dimensions transverse to the $5$-brane and trivial in the other
dimensions.  Thus their correlation functions must be identical, and the
potential will be equal to that of the $d=6$, $\CN=1$ theory.

Rather than make this more explicit here, we argue that a related
$d=6$, $\CN=1$ Lagrangian will also describe the instanton moduli space.
First, it is worth noting that the whole set
up can be considerably generalized in the type \II\ theories, which
contain all even (\IIa) or odd (\IIb) $p$-branes.  By bringing $k$ branes
together we get a world-volume theory with $U(k)$ symmetry, and now
\gsource\ tells us that an instanton in this theory has the right charge to be
a Dirichlet $p-4$-brane.  This suggests that the dimension-dependent details in
the above discussions must not be essential to the basic phenomenon, that
a D-brane can be equivalent to an instanton.

This also suggests that we do the comparison
of the moduli space with the ADHM moduli space in the highest dimension
possible, which will give the strongest constraints on the world-volume
theory.
Let us consider the type \I\ $Sp(N)$ and $U(N)$ theories.
Although choices other than $SO(32)$ are not completely sensible
string theories, the problems come in at one open-string loop (anomalies)
or equivalently closed string tree level (tadpoles).
The classical open string sector will be a sensible
$6$-dimensional supersymmetric effective theory, which we can use
to determine the moduli spaces for the $5$-branes.
This moduli space will then generalize to gauge fields in any $p$-brane --
the idea is that no matter what dimension brane we consider,
the essential structure of the theory is its content in
the four dimensions transverse to the branes.
Using this, we find that
the $Sp(N)$ $k$-instanton moduli space will be reproduced by an $SO(k)$
gauge theory containing $(k,2N)$ half-hypermultiplets and a symmetric
tensor of $SO(k)$, and the $U(N)$ moduli space will be reproduced by
a $U(k)$ gauge theory containing hypermultiplets in the $(k,N)$, adjoint
and singlet.  Indeed all this fits with the description of the ADHM moduli
space given in section 4 of \christ.

Other details of our explicit example will clearly
generalize.  For example, by adding additional branes,
new open string sectors will produce a
fairly general set of fields on the original $p$-brane, and it must be that
the zero modes of these fields in the instanton background
are reproduced by the open strings to the
lower dimensional brane.  The discussion of the transverse moduli should also
be simpler in a more general context.
We plan to give a more general treatment of these points elsewhere.

Let us mention another aspect of the equivalence between gauge soliton
and D-brane, for concreteness using Witten's original example.~\footnote*{
This idea was developed in
discussion with T. Banks, M. Berkooz and N. Seiberg.}
Since we have two descriptions of the same object,
the two sources
\eqn\tsource{dH^{(3)} = -{1\over 8\pi^2} \tr F^2 - 2 *J^{6}}
are really indistinguishable in string theory.
In ordinary gauge theory, gauge fields with different
instanton number are topologically distinct, so asserting that the limit
$F=0$ is connected to an instanton implies a sort of `topology change'
which deserves better understanding.
Consider an $S^3$ surrounding the
instanton (at large distance), and consider the enclosed
magnetic charge $Q=\int H$.
One contribution to this is the integral of the
Chern-Simons form, which is not gauge invariant.  It measures
the instanton number if we define the gauge field using a single patch,
but by a large gauge transformation on the sphere (singular in the interior)
we can change it by an integer.
Evidently this large gauge transformation in string theory must create
a $B$ field on the sphere to keep $Q$ constant, one which could
be produced by a D-brane source in the interior.
Perhaps this can be extended to all of space-time, and we can think
of the equivalence between instanton and D-brane as implemented by
a large gauge transformation.

Another speculation which this suggests is that in a
higher-dimensional formulation (``M-theory''), the instanton in
$d=10$ (e.g. in the boundary theory of \horava) might be a bound
state of a fundamental $5$-brane with the boundary.

In conclusion,
the equivalence of instantons (and possibly other
gauge topological solitons) with $D$-branes
is a very general feature of superstring theory, which surely is an
important clue to its underlying geometry.

\smallskip
I would like to thank T. Banks, M. Berkooz, R. Leigh,
J. Polchinski and N. Seiberg
for enlightening discussions.

\smallskip
This work was supported in
part by DOE grant DE-FG05-90ER40559, NSF
PHY-9157016 and the A. P. Sloan Foundation.

\listrefs
\end